\documentclass{svjour2}
\usepackage{graphicx} \usepackage{times,amsfonts}
\usepackage{mathrsfs}
\journalname{Foundations of Physics}
\begin{document}
\title{Are Causality Violations Undesirable?
\thanks{In honor of the retirement from Davidson College of Dr. L.
Richardson King, an extraordinary teacher and mathematician. An earlier version (gr-qc/0609054v2) was selected as co-winner of the CTC Essay Prize set by Queen Mary College, University of London. The views expressed in this paper are those of the author and should not be attributed to the International Monetary Fund, its Executive Board, or its management. This paper was not prepared using official resources. Comments are appreciated from anonymous referees and from participants in seminars at the Universidad Nacional Aut\'{o}noma de M\'{e}xico and Davidson College.} }
\author{Hunter Monroe}
\institute{H. Monroe, International Monetary Fund, Washington, DC 20431, USA, Tel.: +1-202-623-8020, Fax: +1-202-623-4343, \email{hmonroefop, huntermonroe.com}.}
\date{Received: date / Accepted: date}
\maketitle
\begin{abstract}
Causality violations are typically seen as unrealistic and undesirable features of a physical model. The following points out three reasons why causality violations, which Bonnor and Steadman identified even in solutions to the Einstein equation referring to ordinary laboratory situations, are not necessarily undesirable. First, a space-time in which every causal curve can be extended into a closed causal curve is singularity free---a necessary property of a globally applicable physical theory. Second, a causality-violating space-time exhibits a nontrivial topology---no closed timelike curve (CTC) can be homotopic among CTCs to a point, or that point would not be causally well behaved---and nontrivial topology has been explored as a model of particles. Finally, if every causal curve in a given space-time passes through an event horizon, a property which can be called ``causal censorship'', then that space-time with event horizons excised would still be causally well behaved.
\keywords{General relativity \and Differential geometry \and Spacetime topology}
\end{abstract}
\section{Introduction\label{introduction}}
Causality violations are typically seen as unrealistic and undesirable features of a physical model. The occurrence of causality violations in Bonnor-Steadman solutions to the Einstein equations referring to laboratory situations, for instance, two spinning balls, suggests that the breakdown of relativity's explanatory power is not limited to black hole and big bang singularities.\footnote{See Bonnor \cite{Bonnor2003} and Bonnor and Steadman \cite{Bonnor2004}\cite{Bonnor2005}.} This essay points out several respects in which causality violations are not necessarily an undesirable feature of a physical theory. First, certain conditions entailing causality violations rule out singularities (Section \ref{wellbehaved}). Second, causality-violating space-times may exhibit rich topological behavior (Section \ref{CTCsandtopology}). Third, causality-violating space-times can nevertheless behave in a ``nearly causal'' manner (Section \ref{NearlyCausal}).
\section{Ruling Out Singularities\label{wellbehaved}}
Three typical assumptions of singularity theorems are the absence of causal violations, a nonzero energy density, and the occurrence of a closed trapped surface or trapped set. A natural question is what violations of these assumptions suffice to rule out singularities, with the focus below on violations of causality assumptions. For instance, any compact space-time contains a closed timelike curve (CTC) (Hawking and Ellis \cite{Hawking-Ellis} Proposition 6.4.2) and is singularity-free (Senovilla \cite{Senovilla} Proposition 3.2).

Another more intuitive condition is that every timelike curve be extendible to form a CTC. This condition obviously rules out inextendible timelike curves as well as the existence of trapped regions such as black holes which can be entered but not exited by timelike curves. Thus, the condition simultaneously violates two of typical assumptions mentioned above on causality and trapped surfaces. If this condition holds, gravitational collapse must fail to create a trapped surface separating space-time into exterior and interior regions, just as the Jordan curve theorem fails on a torus. If particles have a topological nature as discussed in the next section, it may be impossible to circumscribe any massive particle by a space-like surface separating space-time into two disconnected components.

An attraction of such conditions for ruling out singularities is that they are purely topological. Furthermore, showing that a causality-violating space-time is singularity-free also shows that the universal cover, which is causally well behaved, is singularity free (Hawking and Ellis \cite{Hawking-Ellis}).

Several results in the literature provide a rationale for disallowing CTCs in physical theories. Tipler \cite{Tipler:1977eb} showed that the emergence of CTCs from regular initial data leads to a singularity. Krasnikov \cite{Krasnikov:2001ci} argues that human beings cannot create a time machine, which he defines as a causal loop $l$ lying in the future of a region $U$ such that a causal loop in the future of $U$ must exist in any maximal extension of $U$. However, neither of these results applies in a totally vicious space-time with a CTC through each point, as Tipler states explicitly. Another objection to CTCs is that moving a massive particle around a CTC to meet its younger self violates conservation of mass-energy; this can be overcome by focussing on vacuum space-times in which particles have a purely topological nature. A vacuum space-time in which all timelike curves are extendible into CTCs violates simultaneously all three typical assumptions of singularity theorems.
\section{Allowing Rich Topological Behavior\label{CTCsandtopology}} CTCs are required for there to be interesting topological behavior, by well-known theorems. Globally hyperbolicity of a space-time implies that it has the uninteresting topology $\mathbb{R}\times\mathscr{S}$ (Geroch \cite{Geroch})\footnote{Bernal and Sanchez \cite{Bernal} have recently shown that the traditional definition of globally hyperbolic can be simplified.} and that causal curves cannot probe the topology of $\mathscr{S}$ by the topological censorship theorem (Friedman \textit{et al} \cite{Friedman:1993ty}). Topology change is potentially desirable, for instance, to model the creation and destruction of particles geometrically in a physical theory based upon empty curved space (Misner and Wheeler \cite{Misner:1957mt}). A useful theory must rule out some phenomena while permitting others, but in fact any two compact 3-manifolds are Lorentz cobordant regardless of topology (Lickorish \cite{Lickorish}). A ``topology selection rule'' is needed, such as the result of Gibbons and Hawking \cite{Gibbons:1991tp}\cite{Gibbons:1992he} that the existence of a spinor structure implies wormholes can be created and destroyed only in multiples of 2. However, no convincing way has been found to relax assumptions to maintain causality and to permit topology change, but not arbitrary topology change (Borde \cite{Borde:1994tx}). This provides a rationale for relaxing causality assumptions.

In addition, causality violations imply a nontrivial topology in that no CTC can be shrunk to a point, as formalized below. Following the literature (for instance Avez \cite{Avez} and Galloway \cite{Galloway198409}), say that two CTCs $\gamma_1$ and $\gamma_2$ are \textit{(freely) timelike homotopic} if there is a homotopy without base-point which continuously deforms $\gamma_1$ into $\gamma_2$ via CTCs; $\gamma_2$ may also be a point rather than a CTC.
Say that a Lorentzian manifold which contains a CTC not timelike homotopic to any point is \textit{timelike multiply connected}. It is known among some practitioners as a folk theorem that any Lorentzian manifold containing a CTC is timelike multiply connected:
\begin{theorem}\label{allwormhole}
No CTC on a Lorentzian manifold $\mathscr{M}$ is timelike homotopic to a point; any $\mathscr{M}$ containing a CTC is timelike multiply connected.
\end{theorem}
\begin{proof}
Suppose there is a CTC $\gamma$ which is timelike homotopic to a point $p$ via the timelike homotopy $F:I\times I\rightarrow \mathscr{M}$, where the curve $F(t,s)$ for fixed $s$ is timelike, $F(0,s)=F(1,s)$ (the timelike curve is closed), and $s$ parameterizes the homotopy from $F(t,0)=\gamma(t)$ to $F(t,1)=p$. Every neighborhood $\mathscr{U}$ of $p$ contains the CTC $F(t,1-\epsilon)$ for sufficiently small $\epsilon>0$. However, Lorentzian manifolds are locally causally well behaved at any point (Garcia-Parrado and Senovilla Proposition 2.1 \cite{Garcia-Parrado:2005yz}).
\end{proof}
Consider two examples of the above theorem.
\begin{example}\label{torus}
Construct a torus by taking the unit square in a 2-D Minkowski space, and identify $t=0$ with $t=1$ and $x=0$ with $x=1$.
\end{example}
In this case, the conclusion of the Theorem \ref{allwormhole} is obvious; the torus is multiply connected, so it is also timelike multiply connected with respect to the strict subset of all homotopies which are timelike.
However, it also follows immediately from Theorem \ref{allwormhole} that a compact simply connected Lorentzian manifold is nevertheless, because it contains a CTC, timelike multiply connected. A contrived example of a manifold which is simply connected (by any type of curve) but timelike multiply connected can be found in Flaherty \cite{Flaherty197503}. However, by Theorem \ref{allwormhole}, the G\"{o}del space-time is also an example, as it contains CTCs and is based upon the simply connected manifold $\mathbb{R}^4$. The following example suggests what prevents a CTC from being contracted to a point on a simply-connected manifold:
\begin{example}\label{S3example}
The 3-sphere $S^3$, which is compact, is simply connected but by Theorem \ref{allwormhole} is timelike multiply connected. Embed $S^3$ in Euclidean space $\mathbb{R}^4$ as $x_0^2+x_1^2+x_2^2+x_3^2=1$. The continuous non-vanishing vector field  $V=(x_1,-x_0,x_3,-x_2)$ defines a Lorentzian metric such that $V$ is everywhere timelike. Consider the CTC $(r\cos \theta, r \sin \theta ,0,\sqrt{1-r^2})$ with parameter $\theta$ for fixed $r$. Contracting the curve by lowering $r$ below $1$, it becomes a null curve at $r=\sqrt{2}/2$ and then becomes a spacelike curve as $r$ is contracted further until it reaches the point $(0,0,0,1)$ when $r=0$ .
\end{example}
This example follows from the Hopf fibration which has fibre $S^1$; Theorem \ref{allwormhole}, which says that no CTC is null timelike homotopic, parallels the result that the Hopf map $\eta:S^3\rightarrow S^2$ is not null homotopic.

Example \ref{S3example} suggests that a CTC cannot be timelike homotopic to a point for one of two reasons: (1) the CTC is not homotopic to a point even allowing homotopies which are not timelike; (2) the CTC is homotopic to a point, but under every such homotopy which is initially timelike, the curve as it is deformed becomes null at some point and the homotopy is no longer a timelike homotopy.

A taxonomy of CTCs could distinguish those which are homotopic to a point and those which are not. Within each category, the set of CTCs can be further divided into equivalence classes under timelike homotopy. These equivalence classes are the elements of the fundamental group under timelike homotopy described by Smith \cite{Smith}.
\section{Behaving in a Nearly Causal Manner}\label{NearlyCausal} This section considers one respect in which a causality-violating space-time can be seen as nearly causal.
Consider which subsets $\mathscr{S}$ of $\mathscr{M}$ have the property that $\mathscr{M}\backslash\mathscr{S}$ contains no closed causal curves, where the set $\mathscr{S}$ is well behaved in some sense, for instance, it is compact, locally spacelike, and edge free. This is the case for the curve $x=1/2$ in Example \ref{torus}, which is a genus-reducing cut that slices open the handle. One may also ask whether there exists a physically significant $\mathscr{S}$, for instance, the set of points at which neighboring geodesics orthogonal to $\mathscr{S}$ are neither converging nor diverging (a condition which holds at $r=2M$ in Schwarzschild coordinates). Such a surface would provide ``causal censorship'' by placing causality violations behind horizons where they cannot be seen from a region of space-time that appears causal to an observer. Thus, excluding causality violations from physical theories simply because we do not see them in practice is risky; a more sophisticated analysis is needed.

In any case, the standard tools of causal analysis can be applied to $\mathscr{M}_c=\mathscr{M}\backslash\mathscr{S}$. In particular, the definitions and results of Hawking and Ellis \cite{Hawking-Ellis} Chapter 6 can be applied to $\mathscr{M}_c$, and $\mathscr{M}_c$ may be causally well-behaved. For instance, $\mathscr{M}_c$ may be globally hyperbolic, with a global Cauchy surface through which all timelike curves in $\mathscr{M}_c$ pass one and only once. Unfortunately, sufficient conditions for causal censorship to hold are not known.
\section{Conclusion\label{Conclusion}}
Hawking \cite{Hawking:1996jh} notes that strong causality assumptions risk ``ruling out something that gravity is trying to tell us'' and that it would be preferable to deduce that causality conditions hold in some region of a given space-time, for instance, $\mathscr{M}_c$. Thus, causality assumptions, like Euclid's parallel postulate, risk closing off interesting lines of investigation.

\end{document}